\documentclass[10pt,twocolumn]{IEEEtran}
\usepackage{graphicx} 
\usepackage{epsfig} 
\usepackage{mathptmx} 
\usepackage{times} 
\usepackage{amsmath} 
\usepackage{amssymb}  
\usepackage{color}

\usepackage[caption=false,font=footnotesize]{subfig}

\DeclareMathOperator*{\argmax}{argmax}

\def\proof{\noindent{{\bf Proof: }}}
\def\endproof{\hspace*{\fill}~\QED\par\endtrivlist\unskip}
\newtheorem{theorem}{Theorem}
\newtheorem{lemma}{Lemma}

\newcommand{\E}[1]{{\mathbb{E}}\left[{#1}\right]}

\newcommand{\Prob}[1]{{\mathbb{P}}\left({#1}\right)}

\newcommand{\Wv}{{\bf W}}
\newcommand{\Rv}{{\bf R}}

\begin{document}

\title{{\LARGE \bf{Wireless Network Control with Privacy
Using Hybrid ARQ}}}

\author{Yunus Sarikaya, Ozgur Ercetin, C.~Emre Koksal
\thanks{ This work was supported in part by European
Commission under Marie Curie IRSES grant PIRSES-GA-2010-269132
AGILENet.} \thanks{ Y.Sarikaya (email: sarikaya@su.sabanciuniv.edu)
and O. Ercetin (email: oercetin@sabanciuniv.edu) are with the
Department of Electronics Engineering, Faculty of Engineering and
Natural Sciences, Sabanci University, 34956 Istanbul, Turkey.}
\thanks{C.~E. Koksal (koksal@ece.osu.edu) is with the Department of
Electrical and Computer Engineering at The Ohio State University,
Columbus, OH.}}
 \maketitle

\begin{abstract}
We consider the problem of resource allocation in a wireless
cellular network, in which nodes have both open and private
information to be transmitted to the base station over block fading
uplink channels. We develop a cross-layer solution, based on hybrid
ARQ transmission with incremental redundancy. We provide a scheme
that combines power control, flow control, and scheduling in order
to maximize a global utility function, subject to the stability of
the data queues, an average power constraint, and a constraint on
the privacy outage probability. Our scheme is based on the
assumption that each node has an estimate of its uplink channel gain
at each block, while only the distribution of the cross channel
gains is available. We prove that our scheme achieves a utility,
arbitrarily close to the maximum achievable utility given the
available channel state information.

\end{abstract}


\section{Introduction}
\label{intro}

Recently, information theoretic security has gained significant
attention, provisioning an ultimate goal of guaranteed security
against adversaries with unlimited computational resources. The
foundations of physical layer secrecy have been initially developed
in~\cite{Wyner,Csiszar} and different variants of the problem
-mainly for the wireless channel- have been revisited vastly. For
example, in~\cite{Lai} channel fading has been exploited for
secrecy, and more recently, multiple antennas~\cite{Khisti} and
cooperative relays~\cite{Dong} have been utilized to increase the
achievable secrecy rates. Despite the significant volume of work in
information theoretic secrecy, most of work has focused on physical
layer techniques and on a single link. The area of wireless
theoretic secrecy remains in its infancy, especially as it relates
to the design of wireless networks and its impact on network control
and protocol development.

To that end, we investigated~\cite{Koksal} the cross-layer resource
allocation problem with information theoretic security. There, we
considered a system in which nodes collect both open and private
information, store them in separate queues and transmit them to the
base station over block fading uplink channels. We first introduced
the concept of private opportunistic scheduling and showed that it
achieves the maximum sum private rate achievable in uplink wireless
systems. We subsequently developed a joint flow control and
scheduling scheme and showed that it achieves a performance, close
to the optimal. In~\cite{Koksal}, we assumed a constant power
transmission and that information is encoded over individual
packets. In this paper, we extend our results to the scenario with
hybrid ARQ (HARQ) transmission based on incremental redundancy
(INR), which basically relies on mutual information accumulation at
each retransmission. Furthermore, we include the possibility that
nodes transmit at varying power levels, subject to an average power
constraint. We assume that the transmitter has an estimate of its
uplink channel and only the distribution of the cross channels to
the every other node. We develop a dynamic cross-layer control
scheme which maximizes aggregate utility subject to power and
privacy constraints. We prove the optimality of our scheme by
Lyapunov optimization theory. Finally, we numerically characterize
the performance of the dynamic control algorithm with respect to
several network parameters.

The HARQ transmission scheme we use is similar to the one employed
in~\cite{X_Tang}, which considers a block fading wire-tap channel
with a single source-destination pair and an (external) eavesdropper
and
develops sequences of Wyner codes to be transmitted in subsequent
transmissions of a block of information. The main challenge of
incorporating information encoding across many blocks into our
solution was that, it is not possible to dynamically update the
resource allocation, based on the amount of information leakage to
the other nodes at each retransmission, since the amount of leakage
is unknown to the transmitting node. Furthermore, the privacy outage
probability of subsequent retransmissions of a given block cannot be
decoupled from each other, eliminating the possibility of using
standard Lyapunov techniques. We resolve that issue by utilizing the
Markov inequality so that the desired decoupling occurs at the
expense of some loss in performance. We believe our new technique
contributes to the field of network control~\cite{Tassiulas,X_Lin},
since it enables the use of Lyapunov techniques in the analysis of
the schemes such as HARQ, which is based on encoding information
over many blocks.

The rest of the paper is organized as follows. Section~\ref{sys_mod}
describes the system model and we provide a brief summary of INR
HARQ. Section~\ref{prob_form} gives the problem formulation. In
Section~\ref{dynamic_control}, we give our joint flow and scheduling
algorithm. Lastly, Section~\ref{numerical} contains the numerical
results of the effect of system parameters on the performance of the
algorithm. Section~\ref{conclusion} concludes this work by
summarizing the contributions.

\section{System Model}
\label{sys_mod}
\begin{figure}[htp]
\begin{center}
  \includegraphics[width=2.0in]{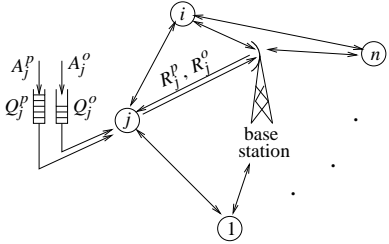}%
\vspace{-0.1in} \caption{Multiuser uplink communication system}
\label{fig:1}       
\end{center}
\end{figure}

\noindent {\bf Network Model:} We consider a multiuser uplink
network as illustrated in Fig. 1. The system consists of $n$ nodes
and a base station. The traffic injected by each of these nodes,
consists of both open and private packets. Nodes wish to transmit
those packets to the base station via the uplink channel, which we
will refer to as the main channel.  All private messages of each
node need to be transmitted to the base station, privately from the
other nodes. They overhear transmissions from the transmitting node
over the cross channels. Hence, nodes will treat each other as
``internal eavesdroppers'' when transmitting private information.

We assume the time to be slotted. Over each block (of time), the
amount of open information, $A_j^o(k)$, and private information,
$A_j^p(k)$ injected in the queues at node $j$ are both chosen by
node $j$ at the beginning of each block. Private and open packets
have a fixed size $\hat{R}_j^p$ and $\hat{R}_j^o$ respectively. Open
and private information are stored in separate queues with sizes
$Q_j^o(k)$ and $Q_j^p(k)$ respectively. At any given block, only one
node transmits either open or private information (but not both) and
a scheduler decides on which node will transmit. In addition, the
scheduler decides the transmission powers for open and private
transmission, which are denoted by $P_j(k)$ and $P_j^o(k)$ for user
$j$'s private and open transmissions, respectively. We use indicator
variables, ${\cal I}_j^o(k)$ and ${\cal I}_j^p(k)$, which take on a
value $0$ or $1$, depending on whether or not open or private
information is transmitted by node $j$ over block $k$.

\noindent {\bf Channel Model:} We assume the block length to be
identical to $N$ channel uses. Both the main and the cross channels
experience independent identically distributed (i.i.d) block fading,
in which the channel gain is constant over a block and it is varying
independently from block to block. We denote the instantaneous
achievable rate for the main channel of node $j$ and the cross
channel between nodes $j$ and $i$ by $R_j(k)$ and $R_{ji}(k)$
respectively. Even though our results are general for all channel
state distributions, in numerical evaluations, we assume all
channels to be Rayleigh fading. Let $h_j(k)$ and $h_{ji}(k)$ be
power gains of the main channel for node $j$ and the cross channel
between node $j$ and node $i$, respectively. We normalize the power
gains such that the additive Gaussian noise has unit variance. Then,
the instantaneous achievable rates are, \vspace{-0.1in}
\begin{align}
R_j(k) &= \log\left(1+P_j(k)h_j(k) \right) \label{main_channel}\\
R_{ji}(k) &= \log\left(1+P_j(k)h_{ji}(k)\right) .
\label{cross_channel} \vspace{-0.2in}
\end{align}
Similarly, the instantaneous achievable rate for the uplink channel
of node $j$ for open messages, $R_j^o(k)$ is:
\begin{equation}
R_j^o(k) = \log\left(1+P_j^o(k)h_j(k) \right) .
\label{main_channel_open}
\end{equation}
We assume that the transmitter has access to a noisy estimate of its
main channel gain and merely the distribution of its cross-channel
gains. After each transmission, the base station informs the
transmitting node about the amount of mutual information accumulated
over that block, i.e., $R_j(k)$ or $R_j^o(k)$.

\noindent {\bf Coding:} We assume that a fixed INR HARQ scheme is
employed at each node. We first explain the details of the version
of the scheme for private packets: Node $j$ collects each packet of
$\hat{R}_j^p$ bits\footnote{Note that, if ${\cal I}_j^p(k)=1$ and
the size of the private queue is smaller than $\hat{R}_j^p$, then
the transmitter uses dummy bits to complete it to $\hat{R}_j^p$.}
and encodes it into a codeword $x_j^{MN}$ called the mother code,
which is then divided into individual groups symbols,
$[x_1^N,x_2^N,\ldots ,x_M^N]$, of length $N$ channel uses. The
mother code is encoded by using Wyner code of $C(\hat{R}_j/M,
\hat{R}_j^p/M,MN)$. After the partitioning is realized, the first
transmission of the packet forms a codeword of Wyner code
$C(\hat{R}_j, \hat{R}_j^p, N)$. At the possible $m^{th}$
transmission, the combined codewords, $[x_1^N,x_2^N,\ldots, x_m^N]$
form a codeword of length $mN$ as $C\left(\hat{R}_j/m,
\hat{R}_j^p/m, mN \right)$. The maximum number of retransmissions is
$M$ and we assume that $M$ is sufficiently large to keep the
probability of decoding failure due to exceeding the maximum number
of retransmissions approximately identical to zero. At each
retransmission, base station combines the codeword of length $N$
with the previously transmitted codewords of the same packet. For a
packet with content $W_j$, let the vector of symbols received by
node $i\neq j$ be $\bold{Y}_i = [Y_{i,1}^N, ...,Y_{i,m}^N]$ at the
end of $m^{th}$ retransmission of the packet by node $j$. To achieve
perfect privacy, following constraint must satisfied by node $j$,
for all $i \neq j$. \vspace{-0.1in}
\begin{align}
\frac{1}{mN}I(W_j;\bold{Y}_i)\leq \epsilon , \label{privacy_const}
\end{align}
for all $\varepsilon >0$. Note that the amount, $\hat{R}_j^p$, of
encoded private information and the amount, $\hat{R}_j$ of bits that
encapsulate the private information is fixed and do not change from
one packet to another. For INR, the mutual information accumulation
for $l^\text{th}$ private packet in the main channel and
eavesdropper channels over block $n$ can be found (as detailed
in~\cite{X_Tang}) as: \vspace{-0.1in}
\begin{align}
D_{j}^l(n) &= \sum_{k = n_j^{l-1}+1}^n \log\left(1 + P_j(k)h_j(k)
{\cal
I}_j^p(k) \right) \\
D_{ji}^l(n) & =  \sum_{k = n_j^{l-1}+1}^n \log\left(1 + P_j(k)
h_{j}(k) {\cal I}_j^p(k) \right)
\end{align}
respectively, where $n_j^{l-1}$ is the block index at which
$(l-1)^\text{st}$ private packet is successfully decoded by the base
station. Note that, if $\hat{R}_j < D_j^l(n)$ at block $n$, we say
that the successful decoding of the private packet took place.

If the accumulated information at one of eavesdroppers exceeds
$\hat{R}_j -\hat{R}_j^p$, perfect privacy constraint
(\ref{privacy_const}) is violated and we say that the privacy outage
occurs. Then, the privacy outage probability over block $n$ for
$l^\text{th}$ private packet is calculated as: \vspace{-0.1in}
\begin{align}
\rho_j^{p,l}(n) = \Prob{\hat{R}_j -\hat{R}_j^p < \max_{i \neq j}
D_{ji}^l(n)}. \label{sec_outage_prob}
\end{align}

For the case of open transmission, at the transmitter, the
information and CRC bits are encoded by a mother
code~\cite{Rowitch}. In each transmission, only the systematic part
of the codeword and a selected number of parity bits are
transmitted. Decoding is attempted at the receiver side by combining
all previously transmitted codes. This procedure is again called INR
HARQ, and mutual information accumulated for $l^\text{th}$ private
packet in the main channel of user $j$ over block $n$ is
\vspace{-0.1in}
\begin{equation}
D_{j}^{l,o}(n) = \sum_{k = n_j^{l_o-1}+1}^n \log\left(1 +
P_j^o(k)h_j(k) {\cal I}_j^o(k) \right), \vspace{-0.1in}
\end{equation}
where $n_j^{l_o-1}$ corresponds to the block index, where
$(l-1)^\text{st}$ open message is successfully decoded by the base
station. Here, we assume that fixed length packets are encoded with
a rate of $\hat{R}_j^o$, arrive to the open queue at node $j$. If
the accumulated information is larger than the fixed rate, i.e.,
$\hat{R}_j^o < D_j^{l,o}(n)$, the decoding of the open message is
successful.

\section{Problem Formulation}
\label{prob_form}

In this section, we formulate the problem as a network utility
maximization (NUM) problem. Our objective is to choose the admission
rate and transmission power in order to achieve a long term private
and open rates close to the optimal, while keeping the rate of
privacy outages below a certain level.

Let $U_j^p(x)$ and $U_j^o(x)$ be the utilities obtained by node $j$
from the transmission of $x$ private and open bits respectively. We
assume that $U_j^p(\cdot)$ and $U_j^o(\cdot)$ are non-decreasing
concave functions and the utility of a private information is higher
than the utility of open transmission at the same rate, i.e.,
$U_j^p(x)>U_j^o(x)$. In addition, it is assumed that the arrival
processes are ergodic.

To state the problem clearly, we define the expected service and the
expected arrival rates of the private and open queues at each node
as follows. First, the amount of private information transmitted
from node $j$ in block $k$ is $R_j^p(k) \triangleq
\frac{\hat{R}_j^p}{\hat{R}_j}R_j(k)$, since
$\frac{\hat{R}_j^p}{\hat{R}_j}$ is the fraction of the private
information encapsulated within $R_j(k)$ bits of transmitted data.
Let $\mu_j^p$ and $\mu_j^{o}$ denote the expected service rates of
private and open traffic queues, respectively, i.e.,
$\mu_j^p=\E{{\cal I}_j^p(k)R_j^p(k)}$ and $\mu_j^o=\E{{\cal
I}_j^o(k)R_j^o(k)}$.  Note that, the ``effective'' expected rate of
private information received at the base station without a privacy
outage is $\mu_j^{p,e}$. Hence, node $j$ effectively obtains an
utility of $U_j^p(\mu_j^{p,e})$ from its private transmissions. We
assume that the utility gained by a packet suffering a privacy
outage reduces from that of a private packet to that of an open
packet. Thus, node $j$ obtains an utility of
$U_j^o(x_j^p-\mu_j^{p,e}+x_j^o)$ from all transmitted open messages
as well as the messages that have been encoded privately, but have
undergone a privacy outage. Finally, let the expected arrival rate
to the private queue of node $j$ be $x_j^p\triangleq \E{A_j^p(k)}$
and the expected arrival rate to the open queue of node $j$ be
$x_j^o\triangleq \E{A_j^o(k)}$. We consider the following
optimization problem:

\vspace{-0.2in} \small
\begin{align}
\max
\sum_{j=1}^n&\left(
U_j^p(\mu_j^{p,e})+U_j^o(x_j^p-\mu_j^{p,e}+x_j^o))\right) \\
\mbox{ subject to}\quad &x_j^p \leq \mu_j^p , \forall j,\
x_j^o \leq \mu_j^o , \forall j \label{queue_constraint}\\
& \mu_j^{p,e}/\mu_j^p \geq 1- \gamma_j \label{secrecy_outage_constraint} \\
& \E{P_j(k) + P_j^o(k)} \leq \alpha_j , \forall j,
\label{power_constraint}
\end{align}
\normalsize where the maximization is over the parameters $\{{\cal
I}_j^p(k),{\cal I}_j^o(k),P_j(k),P_j^o(k),A_j^p(k),A_j^o(k)\}$; the
constraints in (\ref{queue_constraint}) ensure the stability of
private and open queues, respectively;
(\ref{secrecy_outage_constraint}) corresponds to the privacy outage
constraint, which ensures that portion of private packets
intercepted by the eavesdroppers is below of some threshold,
$\gamma_j$, and (\ref{power_constraint}) correspond to average power
constraint.

The challenge in our problem lies in the fact that, the objective
functions of the nodes are coupled. In other words, the private
utility function of each node depends on scheduling decision, which
inevitably affects the utilities obtained by all nodes in the
system. In order to decouple utilities obtained by each user from
their private transmissions, we introduce an auxiliary variable
$x_j^{p,e}$ for each variable $\mu_j^{p,e}$. By introducing
auxiliary variables, we add a new set of constraints and the
optimization problem becomes:

\vspace{-0.2in} \small
\begin{align}
\max \sum_{j=1}^n&\left(
U_j^p(x_j^{p,e})+U_j^o(x_j^p-x_j^{p,e}+x_j^o))\right) \label{objective_function} \\
\mbox{ subject to } \quad &x_j^p \leq \mu_j^p , \forall j,\
x_j^o \leq \mu_j^o , \forall j \label{queue_constraint1}\\
&x_j^{p,e} \leq \mu_j^{p,e} , \forall j  \label{eff_private_queue_constraint}\\
&x_j^{p,e}/x_j^p \geq 1-\gamma_j , \forall j
\label{secrecy_outage_constraint1}\\
& \E{P_j(k) + P_j^o(k)} \leq \alpha_j  , \forall j.
\label{power_constraint1}
\end{align}
\normalsize

Note that $x_j^{p,e}$ in (\ref{eff_private_queue_constraint}) can be
interpreted as the long term average arrival rate for packets which
do not incur privacy outage. Thus, the portion of packets, kept
private from eavesdroppers should be greater than $1-\gamma_j$. Also
note that since objective function is an increasing function of
$x_j^{p,e}$ (\ref{eff_private_queue_constraint}) is satisfied with
equality at the optimal point.


\section{Dynamic Control}
\label{dynamic_control}

In this section, we present an opportunistic scheduling algorithm
maximizing the total expected utility of network while satisfying
the constraints \eqref{queue_constraint1}-\eqref{power_constraint1}.
In the following, we assume that there is an infinite backlog of
data at the transport layer of each node providing both private and
open messages. The private messages are encoded by Wyner code at a
fixed rate $(\hat{R}_j,\hat{R}_j^p)$.

However, the challenge here is that the privacy outage probability
in (\ref{sec_outage_prob}) depends on the past transmissions and the
scheduling decision may affect future transmissions, i.e., the
events that successful decoding occurs by an eavesdropper over
subsequent retransmissions are non-iid. Then, utilizing standard
Lyapunov optimization techniques~\cite{neely} to solve our problem
is not possible. To address this issue, we need to quantify the
privacy outage probability over each block independently. For that
purpose, we make use of Markov's inequality: \vspace{-0.1in}

\vspace{-0.1in} \footnotesize
\begin{align}
    &\Prob{\hat{R}_j - \hat{R}_j^p < \max_{i\neq j} D_{ji}^l (n)} \leq \frac{1}{\hat{R}_j-\hat{R}_j^p}\left(1-
    \prod_{i\neq j}(1-\E{D_{ji}^l (n)})\right) \label{Markov_inequality1} \\
     &= \frac{1}{\hat{R}_j-\hat{R}_j^p} \left(\sum_{i \neq j} \E{D_{ji}^l (n)} \prod_{h>j, h\neq j} (1-\E{D_{jh}^l (n))}) \right) \label{Markov_inequality2} \\
    &\leq \sum_{i \neq j} \frac{1}{\hat{R}_j-\hat{R}_j^p}\E{D_{ji}^l(n)} = \frac{1}{\hat{R}_j-\hat{R}_j^p} \sum_{k = n_j^{h-1}+1}^n \sum_{i \neq j}  \E{ {\cal I}_j^p(k) R_{ji}(k)}, \label{Markov_inequality3}
\end{align}
\normalsize where (\ref{Markov_inequality1}) follows from  Markov
inequality, and (\ref{Markov_inequality2}) is due to the fact that
we choose the pair $(\hat{R}_j,\hat{R}_j^p)$ such that
$\E{D_{ji}^l(n)} < \hat{R}_j-\hat{R}_j^p$ for all $j$. Recall that
$D_{ji}^l (n)$ is the accumulated information at the eavesdropper
$i$. According to the Markov inequality, the fraction of private
packets that suffer a privacy outage is thus upper bounded by
$\frac{1}{\hat{R}_j-\hat{R}_j^p}\sum_{i \neq j} \E{R_{ji}(k)}$ for
any block $k$. Hence, Markov inequality enables us to quantify the
amount of information leakage to the other nodes independently over
each block $k$, being
$\frac{\hat{R}_j^p}{\hat{R}_j-\hat{R}_j^p}\sum_{i \neq j}
\E{R_{ji}(k)}$. However, since the Markov inequality is merely a
bound, the constraint set over which we solve the problem shrinks.
Hence some performance is sacrificed. In the simulations, we
numerically analyze the amount of shrinkage in the constraint set
due to the use of the Markov inequality and show that it is not
significant under most scenarios.

The dynamics of private and open traffic queues, $Q_j^p(k)$ and
$Q_j^o(k)$ respectively, are given as follows:

\vspace{-0.2in} \small
\begin{align}
Q_j^p(k+1)&=\left[ Q_j^p(k)-{\cal I}_j^p(k)R_j^p(k)\right]^+ +
A_j^p(k),\\
Q_j^{o}(k+1)&=\left[ Q_j^{o}(k)-{\cal I}_j^o(k) R_j^o(k)\right]^+ +
A_j^o(k),
\end{align}
\normalsize where $[x]^+ = \max(0,x)$.

As shown in \cite{neely}, each of the constraints
\eqref{eff_private_queue_constraint}-\eqref{power_constraint1} can
be represented by a virtual queue, and when these virtual queues are
stabilized the constraints are also satisfied.

\vspace{-0.2in} \small
\begin{align}
Q_j^{p,e}(k+1)&=\left[ Q_j^{p,e}(k)-{\cal I}_j^p(k)
R_j^{p,e}(k)\right]^+ + {\cal A}_j^{p,e}(k), \label{virtuel_queue1}\\
Z_j(k+1) & = \left[ Z_j(k) -{\cal A}_j^{p,e}(k)+
A_j^{p}(k)(1-\gamma_j)\right]^+ \label{virtuel_queue2}\\
Y_j(k+1)&= \left [Y_j(k) + {\cal I}_j^p(k)P_j(k) + {\cal
I}_j^p(k)P_j^o(k) -\alpha_j \right]^+ \label{virtuel_queue3}
\end{align}
\normalsize where virtual queues in
(\ref{virtuel_queue1}-\ref{virtuel_queue3}) represent the
constraints in
(\ref{eff_private_queue_constraint}-\ref{power_constraint1})
respectively. In addition, $R_j^{p,e}(k)$ denotes the private
information sent to the base station without privacy outage over
block $k$. By using the result of Markov inequality in
(\ref{Markov_inequality3}), we obtain $R_j^{p,e}(k)$ as $R_j^p(k)  -
\frac{\hat{R}_j^p}{\hat{R}_j-\hat{R}_j^p}\sum_{i \neq j}
\E{R_{ji}(k)}$. The first term corresponds to the amount of
information received by the base station and the second term to the
amount of information captured by the eavesdroppers.

\noindent {\bf Control Algorithm:} The algorithm executes the
following steps in each block $k$:
\begin{enumerate}
\item[\bf (1)] {\bf Flow control:} For some $V>0$, each
node $j$ injects $A_j^p(k)$, and $A_j^o(k)$ bits to respective
queues and update the virtual queue with ${\cal A}_j^{p,e}$. Note
that ${\cal A}_j^{p,e}(k)$ can be interpreted as private bits for
which perfect secrecy constraint is intended to be satisfied.

\vspace{-0.1in} \small
\begin{align*}
&\left( A_j^p(k),{\cal
A}_j^{p,e}(k),A_j^o(k) \right) = \\
&\argmax \left\{V\left[
U_j^p({\cal A}_j^{p,e}(k))+U_j^o(A_j^p(k)-{\cal A}_j^{p,e}(k)+A_j^o(k)) \right]\right. \\
&\left. -\left( Q_j^p(k)A_j^p(k)+Q_j^o(k)A_j^o(k)+Z_j(k)(
A_j^{p}(k)(1-\gamma_j)-{\cal A}_j^{p,e}(k)) \right)\right\}
\end{align*}
\normalsize
\item[\bf (2)] {\bf Scheduling:} At any given block, scheduler chooses which node will transmit and
the amount of power used during transmission of private messages. In
other words, schedule node $j$ and transmit privately encoded
(${\cal I}_j^p=1$), or open bits (${\cal I}_j^{o}=1$), with transmit
power $P_j$ and $P_j^o$:

\vspace{-0.1in} \small
\begin{align}
&\left({\cal I}_j^p(k),{\cal I}_j^{o}(k),P_j(k),P_j^o(k)\right) = \nonumber \\
&\argmax \left\{ {\cal I}_j^p(k)Q_j^{p,e}(k)\E{R_j^{p,e}(k)}+{\cal
I}_j^p(k)Q_j^p(k)\E{R_j^p(k)} \right.\nonumber\\
&\left.+Q_j^o(k)\E{R_j^o(k)}- Y_j(k)({\cal I}_j^p(k)P_j(k) + {\cal
I}_j^o(k)P_j^o(k) \right\},\nonumber
\end{align}
\normalsize where expectation is over the distribution of channel
estimation error over block $k$.
 \vspace{0.0in}

\end{enumerate}
\normalsize

\subsubsection{Optimality of Control Algorithm}
\label{Optimality}

The optimality of the algorithm can be shown using the Lyapunov
optimization theorem. Let $\mathbf{Q^p(k)}=(Q^p_1(k),\ldots,
Q^p_n(k))$, $\mathbf{Q^o(k)}=(Q^o_1(k),\ldots, Q^o_n(k))$,
$\mathbf{Q^{p,e}(k)}=(Q^{p,e}_1(k),\ldots, Q^{p,e}_n(k))$,
$\mathbf{Z(k)}=(Z_1(k),\ldots, Z_n(k))$,
$\mathbf{Y(k)}=(Y_1(k),\ldots, Y_n(k))$ be the vectors of real and
virtual queues. We consider a quadratic Lyapunov function of the
form:

\vspace{-0.15in} \footnotesize
\begin{align}
L(k) = \frac{1}{2}\sum_j \left[(Q^p_j(k))^2+(Q^o_j(k))^2+
(Q_j^{p,e}(k))^2 + (Z_j(k))^2 + (Y_j(k))^2\right].
\label{eq:lyapunov-function}
\end{align}
\normalsize

One-step expected Lyapunov drift, $\Delta(k)$ is the difference
between the value of Lyapunov function at the $(k+1)^\text{st}$
block and $(k)^\text{th}$ block.

The following lemma provides an upper bound on $\Delta(k)$.
\begin{lemma}
\label{lemma:drift-1} \small
\begin{align}
\Delta(k)&\leq B - \sum_j \E{Q^p_j(k)({\cal
I}_j^p(k)R^p_j(k)- A^p_j(k))|\ Q^p_j(k)} \nonumber  \\
&-\sum_j \E{Q^o_j(k)({\cal I}_j^o(k)R^o_j(k)-A^o_j(k))|\ Q^o_j(k)}
\nonumber \\
&-\sum_j \E{Q^{p,e}_j(k)({\cal I}_j^p(k)R^{p,e}_j(k)-{\cal
A}^{p,e}_j(k))|\
Q^{p,e}_j(k)} \nonumber \\
&-\sum_j \E{Z_j(k)({\cal A}^{p,e}_j(k)- (1-\gamma_j)A^{p}_j(k))|\
Z_j(k)} \nonumber \\
&-\sum_j \E{Y_j(k)(\alpha_j-{\cal I}_j^p(k)P_j(k)-{\cal
I}_j^o(k)P_j^o(k))|\ Y_j(k)} \label{eq:delta}
\end{align}
where $B>0$ is a constant.
\end{lemma}
\proof 
In an interference-limited practical wireless system both the the
transmission power and the transmission rate are bounded. Assume
that the arrival rates are also bounded by $A^{p,max}_j$, ${\cal
A}^{p,e,max}_j$, $A^{o,max}_j$. By simple algebraic manipulation one
can obtain a bound for the difference $(Q^p_j(k +
1))^2-(Q^p_j(k))^2$ and also for other queues to obtain the result
\eqref{eq:delta}.
%
%
\endproof

We now present our main result showing that our proposed dynamic
control algorithm can achieve a performance arbitrarily close to the
solution of the problem with the outage constraint tightened via
Markov's inequality.
\begin{theorem}
\label{thm:optimalcontrol-1}
 If $R_j^p(k) < \infty$ and $R_j^o(k) <\infty$ for all $j,k$, then dynamic control
algorithm satisfies: \vspace{-0.1in}
\begin{align*}
\liminf_{N\rightarrow\infty}\frac{1}{N}\sum_{k=0}^{N-1}\sum_{j=1}^n
\E{U_j^p(k)+U_j^o(k)} &\geqslant U^* - \frac{B}{V} \\
\limsup_{N\rightarrow\infty}\frac{1}{N}\sum_{k=0}^{N-1}\sum_{j=1}^n
\E{Q_j^p(k)} &\leqslant \frac{B+V(\bar{U}-U^*)}{\epsilon_1} \\
\limsup_{N\rightarrow\infty}\frac{1}{N}\sum_{k=0}^{N-1}\sum_{j=1}^n
\E{Q_j^o(k)} &\leqslant \frac{B+V(\bar{U}-U^*)}{\epsilon_2} ,
\end{align*}
where $B,\epsilon_1,\epsilon_2>0$ are constants, $U^*$ is the
optimal aggregate utility and $\bar{U}$ is the maximum possible
aggregate utility.
\end{theorem}

\proof The proof of Theorem \ref{thm:optimalcontrol-1} is given in
Appendix.
\endproof

\section{Numerical Results}
\label{numerical}

In our numerical experiments, we consider a network consisting of
four nodes and a single base station. The main channel between the
node and the base station, and the cross-channels between nodes are
modeled as iid Rayleigh fading Gaussian channels.  The power gains
of the main and cross-channels are exponentially distributed with
means uniformly chosen in the intervals [25,50], [0.5,1.5],
respectively. The main channel power gain is estimated by an
unbiased estimator based on the a priori channel measurements. As
discussed in \cite{Frenger}, the estimation error of such an
estimator, $e_j(k)$ can be modeled with a zero mean finite variance
Gaussian random variable, i.e., $e_{ji}(k) \sim {\cal
N}(0,\sigma^2)$ for all $k$. We take $\sigma= 1$. In addition, we
assume only the knowledge of distribution for the cross-channel
gains.

We consider logarithmic private and open utility functions where the
private utility is $\kappa$ times more than open utility at the same
rate. More specifically, for a scheduled node $j$,
$U_j^p(k)=\kappa\cdot\log(1+R_j^p(k))$, and
$U_j^o(k)=\log(1+R_j^o(k))$.  We take $\kappa=5$ in all experiments.
We perform the simulation over five realizations of $\hat{R}_j$,
$\hat{R}_j^o$ and $\hat{R}_j^p$. $\hat{R}_j$ and $\hat{R}_j^o$ are
uniformly chosen in the interval [15, 25] and $\hat{R}_j^p$ in the
interval [5, 10]. The rates depicted in the graphs are per node
arrival and service rates averaged over all realizations of
$\hat{R}_j$ and $\hat{R}_j^p$ , i.e., the unit of the plotted rates
are bits/channel use/node.  All nodes have the same privacy outage
probability $\gamma$.
\begin{figure}
\centering
\begin{tabular}{c}
\epsfig{file=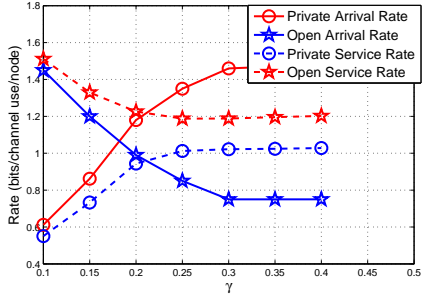,width=2.5in,clip=}\\
(a) Rate vs $\gamma$  \\
\end{tabular}
\begin{tabular}{c}
\epsfig{file=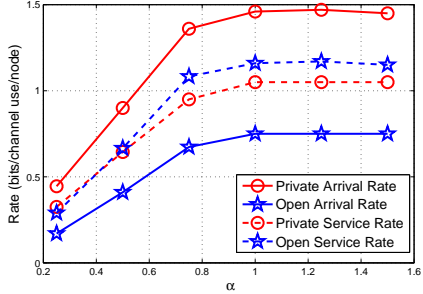,width=2.5in,clip=}\\
(b) Rate vs $\alpha$ \\
\end{tabular}
\caption{Private and open rates with respect to tolerable privacy
outage probability and average power constraint} \label{fig:8}
\end{figure}
In Fig. 2(a), we investigate the effect of the tolerable privacy
outage probability. It is interesting to note that private service
rate increases with increasing tolerable outage probability,
$\gamma$. This is due to the fact that for low $\gamma$ values, the
privacy outage condition is very tight, and this condition is
satisfied by transmitting infrequently only when the channel is at
its best condition and with low transmit power. The highest private
service rate is realized when $\gamma = 0.3$, which suggests that
30\% of the private packets undergo privacy outage. In Fig. 2(b),
the effect of average power constraint, $\alpha$ is investigated. As
expected, for a tight power constraint, all rates are lower, since
selected powers are smaller. The highest rates are obtained when
$\alpha = 1$, and after $\alpha = 1$, the power constraint becomes
inactive. In addition, the bound on privacy outage probability
obtained by Markov inequality is 0.22, which is obtained by
averaging the resulting values of the bound over all simulations,
whereas the privacy outage probability calculated as
in~\eqref{sec_outage_prob} is approximately 0.18. In most of the
scenarios, this difference is not significant as long as privacy
outage constraint is satisfied.

\section{Conclusion}
\label{conclusion}

We consider the problem of resource allocation in a wireless
cellular network, in which nodes have both open and private
information to be transmitted to the base station over block fading
uplink channels. We have developed a cross-layer dynamic control
algorithm in the presence of imperfect knowledge based on hybrid ARQ
transmission with incremental redundancy. We explicitly took into
account the privacy and power constraints and prove the optimality
of our scheme by Lyapunov optimization theory. The main challenge
that we faced is that, due to encoding of information across many
blocks, the privacy outage probability of subsequent retransmissions
of a given block cannot be decoupled from each other. We overcame
this challenge by introducing a novel technique based on the Markov
inequality.


\appendix

Lyapunov Optimization Theorem suggests that a good control strategy
is the one that minimizes the following:

\vspace{-0.2in} \small
\begin{equation} \Delta^U(k)=\Delta(k) - V \E{\sum_j
\left(U^p_j(k)+U^o_j(k)\right) |
\left(\mathbf{Q^p(k)},\mathbf{Q^o(k)}\right)}
\label{eq:deltawithreward}
\end{equation}
\normalsize where $U^p_j(k)$ and $U^o_j(k)$ are private and open
utility obtained in block $k$.

By using \eqref{eq:delta}, we may obtain an upper bound for
\eqref{eq:deltawithreward}, as follows:

\vspace{-0.2in} \small
\begin{align}
&\Delta^U(k)< B-\sum_j \mathbb{E}\left[ Q^p_j(k)[{\cal
I}_j^p(k)R^p_j(k)-A^p_j(k)]|Q^p_j(k)\right]\nonumber\\
&-\sum_j \mathbb{E}\left[ Q^o_j(k)[{\cal
I}_j^o(k)R^o_j(k)-A^o_j(k)]|Q^o_j(k)\right] \nonumber\\
&-\sum_j \mathbb{E}\left[ Q^{p,e}_j(k)[{\cal
I}_j^p(k)R^{p,e}_j(k)-{\cal A}^{p,e}_j(k)]|Q^{p,e}_j(k)\right] \nonumber\\
&-\sum_j \mathbb{E}\left[ Y_j(k)[\alpha_j- {\cal
I}_j^p(k)P_j(k)-{\cal
I}_j^o(k)P_j^o(k)]|Y_j(k)\right] \nonumber\\
&-\sum_j \mathbb{E}\left[ Z_j(k)[{\cal A}^{p,e}_j(k)-(1-\gamma_j)A_j^p(k)]|Z_j(k)\right] \nonumber\\
& -V \E{\sum_j U^p_j({\cal A}_j^{p,e}(k))+\sum_j
U^o_j(A_j^p(k)-{\cal A}_j^{p,e}(k)+A_j^o(k))} \label{drift_final}
\end{align}
\normalsize

Thus, by rearranging the terms in \eqref{drift_final} it is easy to
observe that our proposed dynamic network control algorithm
minimizes the right hand side of \eqref{drift_final} with the
available channel information.

Assume that there exists a stationary scheduling and rate control
policy that chooses the users and their transmission powers
independent of queue backlogs and only with respect to the channel
statistics. Let $U^*$ be optimal value of the objective function of
the problem (\ref{objective_function}) by the stationary policy.
Also let $x_j^{p,e*}$, $x_j^{p*}$ and $x_j^{o*}$ be optimal
effective private, private and open traffic arrivals. In addition,
let $P_j^*$ be optimal transmission power for user $j$. Note that,
the expectations of right hand side (RHS) of \eqref{drift_final} can
be written separately due to independence of backlogs with
scheduling and rate control policy.
Since the rates and transmission power are strictly interior of the
feasible region, the stationary policy should satisfy the following:

\vspace{-0.15in} \footnotesize
\begin{align}
    \mathbb{E}\left[ {\cal I}_j^p(k)R^p_j(k)\right] &\geq x_j^{p*} +
    \epsilon_1 \mbox{ ,  } \mathbb{E}\left[ {\cal I}_j^p(k)R^{o}_j(k)\right] \geq x_j^{o*} +
    \epsilon_2 \nonumber\\
     \mathbb{E}\left[ {\cal I}_j^o(k)R^{p,e}_j(k)\right] &\geq x_j^{p,e*} +
    \epsilon_3 \mbox{  ,  } \mathbb{E}\left[ {\cal I}_j^p(k)P_j(k)-{\cal I}_j^o(k)P_j^o(k)\right] \leq \alpha_j + \epsilon_4 \nonumber\\
     x_j^{p,e*} &\geq (1-\gamma_j)x_j^{p*} +
    \epsilon_5 \label{interior_eq}
\end{align}
\normalsize

Recall that our proposed policy minimizes RHS of
\eqref{drift_final}, hence, any other stationary policy has a higher
RHS value. By using optimal stationary policy, we can obtain an
upper bound for the RHS of our proposed policy. Inserting
\eqref{interior_eq} into \eqref{drift_final} and using the
independence of queue backlogs with scheduling and rate policy, we
obtain the following bound:

\vspace{-0.2in} \small
\begin{align}
RHS<&B-\sum_j\epsilon_1\mathbb{E}\left[Q_j^p(k)\right]-\sum_j\epsilon_2\mathbb{E}\left[Q_j^{o}(k)\right]-\sum_j\epsilon_3\mathbb{E}\left[Q_j^{p,e}(k)\right]
\nonumber \\
&-\sum_j\epsilon_4\mathbb{E}\left[Y_j(k)\right]
-\sum_j\epsilon_5\mathbb{E}\left[Z_j(k)\right] -VU^*.\nonumber
\end{align}
\normalsize

Now, we can obtain bounds on performance of the proposed policy and
the sizes of queue backlogs as given in Theorem 1.

\end{document}